\documentclass[prl,aps,showpacs,twocolumn,preprintnumbers]{revtex4}
\usepackage{amsfonts}
\usepackage{amsmath}
\usepackage{amssymb}
\usepackage{graphicx}
\usepackage{dcolumn}

\def\R{\mathbb R}

\begin{document}

\title{Self-isospectrality, special supersymmetry, and their
              effect on the band structure}
\author{\textsf{Francisco Correa${}^1$, V\'{\i}t Jakubsk\'y${}^1$, Luis-Miguel Nieto${}^2$, and
Mikhail S. Plyushchay${}^1$}
 \\
{\small \textit{${}^1$Departamento de F\'{\i}sica, Universidad de
Santiago de Chile, Casilla 307, Santiago 2, Chile}}\\
{\small \textit{${}^2$Departamento de F\'{\i}sica Te\'orica,
At\'omica y \'Optica, Universidad de Valladolid, 47071,
Valladolid, Spain}} \\
\date{}
}

\begin{abstract}
\noindent We study a planar model of a non-relativistic electron in
periodic magnetic and electric fields that produce a 1D crystal for
two spin components separated by a half-period spacing. We fit the
fields to create a self-isospectral pair of finite-gap associated
Lam\'e equations shifted for a half-period, and show that the system
obtained is characterized by a new type of supersymmetry. It is a
special nonlinear supersymmetry generated by three commuting
integrals of motion, related to the parity-odd operator of the
associated Lax pair, that coherently reflects the band structure and
all its peculiarities. In the infinite period limit it provides an
unusual picture of supersymmetry breaking.
\end{abstract}

 \maketitle


\emph{Introduction -- } Supersymmetry as a fundamental symmetry of
Nature still waits for experimental confirmation,  but as a kind
of symmetry between bosonic and fermionic states, it already
turned out to be fruitful in diverse areas, including nuclear
\cite{IachNuc,NucSUSY}, atomic, solid-state, and statistical
physics \cite{SUSYQM}. Supersymmetric quantum mechanics
\cite{SUSYQM} was introduced under investigation of the problem of
supersymmetry breaking in field theory \cite{Wit}. In the simplest
case a system is characterized there by a $2\times 2$ diagonal
matrix Hamiltonian, $H=diag(H^+,H^-)$, and by two anti-diagonal
matrix integrals of motion (supercharges) $Q_1$ and
$Q_2=i\sigma_3Q_1$. Supercharges are first order differential
operators generating an $N=2$ superalgebra
\begin{equation}\label{susylin}
    \{Q_a,Q_b\}=2\delta_{ab}H, \quad [Q_a,H]=0.
\end{equation}
Such a
system has an additional integral of motion $\Gamma=\sigma_3$,
$\Gamma^2=1$, which classifies the states with $\Gamma=+1$ and $-1$,
by convention, as bosonic and fermionic states. Since $[\Gamma,H]=0$
and $\{\Gamma,Q_a\}=0$, the Hamiltonian and supercharges are
identified as bosonic and fermionic generators. The cases of
unbroken and broken supersymmetry are distinguished by the Witten
index $\Delta$ defined as the difference between the total numbers
of bosonic and fermionic states. In  a \emph{non-periodic}
one-dimensional system, unbroken supersymmetry is characterized by
one singlet ground state of zero energy and $\Delta\neq 0$; for
broken supersymmetry there is no zero energy singlet state and
$\Delta=0$. However, as it was observed in \cite{mac,dunfei},
supersymmetric \emph{periodic} systems may support two zero energy
ground states, and then $\Delta=0$  in the unbroken case. Some
systems investigated there possess a specific property of
\emph{self-isospectrality}, meaning that corresponding superpartner
potentials $V^+$ and $V^-$ are given by the same periodic function
shifted for a half of a period $2L$, $V^+(x+L)=V^-(x)$. They belong
to a class of finite-gap periodic systems, which play an important
role, in particular,  in condensed matter physics \cite{AGI} and in
the theory of nonlinear integrable systems \cite{Int}.

In a \emph{nonlinear} generalization of supersymmetric quantum
mechanics \cite{ais},  supercharges $Q_a$ are higher ($n>1$) order
differential operators generating a nonlinear superalgebra
$\{Q_a,Q_b\}=2\delta_{ab}P_n(H)$, with $P_n(H)$ a polynomial of
order $n$. The number of singlet states can take there any value
from $0$ to $n$, and, as in periodic models with linear
supersymmetry, the Witten index does not characterize supersymmetry
breaking \cite{ais,KP}. This indicates that in periodic finite-gap
systems nonlinear supersymmetry may play an important role.

To investigate the question of the presence and nature of
nonlinear supersymmetry in periodic finite-gap systems, in this
Letter we study a planar model described by the Pauli Hamiltonian
for a non-relativistic electron in periodic electric and magnetic
fields. The model belongs to a broad class of periodic systems
investigated by Novikov \emph{et al.} \cite{Novikov}. It is well
known that \emph{in the absence} of an electric field the model,
which includes the Landau problem as a particular case, is
characterized by a supersymmetry with the usual linear
superalgebraic structure (\ref{susylin})
\cite{SUSYQM}. We choose
periodic magnetic and electric fields in such a form that the
spin-up and -down components of the electron wave function feel
the same one-dimensional effective periodic potential but with a
shift of half of the period. As a result, the effective potential
of superextended system satisfies a property of
self-isospectrality. Vector and scalar potentials are fitted to
produce the associated Lam\'e equation with two integer parameters
$m$ and $l$, which belongs to a broad class of finite-gap systems
with a smooth potential, see Eq. (\ref{V}) below
\cite{Int,Gang,Com1}. We find here a special nonlinear
supersymmetry of the previously unknown structure, in which all
the peculiarities of the band structure of the system are
imprinted. In  the infinite-period limit our system provides an
unusual picture of supersymmetry breaking rooted in its
nonlinearity.

\emph{Model -- } Consider a non-relativistic electron confined to
a plane and moving in the presence of an electric field, given by
a scalar potential $\phi(x,y)$, and a perpendicular magnetic field
$B_z(x,y)$. It is described by the Pauli Hamiltonian
\begin{equation}
    H_e=\left( p_x+A_{x}\right) ^{2}+\left( p_y+A_{y}
    \right) ^{2}+\sigma_{3}B_{z}-\phi,
    \label{Hel}
\end{equation}
where the units are $\hbar =c=2m=-e=1$. Let us restrict  $B_z$ and
$\phi$ by the condition that they depend only on $x$. We choose
$A_{x}=0,$ $A_{y}=w(x)$, then  $B_{z}=\frac{dw}{dx}$.  Present the
wave function in the form $ \Psi (x,y)=e^{i\kappa y}\psi (x)$, where
$\kappa$, $-\infty<\kappa<\infty$, is the eigenvalue of $p_{y}$.
Taking $w(x)=\alpha\frac{d}{dx}\ln (\mathrm{dn}x)$  and
$\phi(x)=\beta w^2(x) +\gamma w(x) +\delta$, with appropriate choice
of constant parameters $\alpha$, $\beta$, $\gamma$ and $\delta$, we
reduce (\ref{Hel}) to a quantum periodic system given by the
diagonal matrix Hamiltonian $H$ with up ($+$) and down ($-$)
components of the form $H^\pm_{m,l}=-\frac{d^2}{dx^2}
+V^\pm_{m,l}(x)$. Here $V_{m,l}^{+}(x)= V_{m,l}^{-}(x+L)$,
\begin{equation}
    V_{m,l}^{-}(x)=-C_m\mathrm{dn}^{2}x-C_l\frac{
    k^{\prime }{}^{2}}{\mathrm{dn}^{2}x}+c,
    \label{V}
\end{equation}
$C_m=m(m+1)$, $C_l=l(l+1)$, $m$ and $l$ are integers such that
$C_m^2+C_l^2\neq 0$, $c$ is a constant;
$\mathrm{dn}\,x=\mathrm{dn}\,(x,k)$ is the Jacobi even elliptic
function, satisfying a relation
$\mathrm{dn}\,(x+K)=k'/{\mathrm{dn}\,x }$, with modulus $0<k<1$
and real and imaginary periods $2K$ and $4iK'$,  $K(k)$ is the
elliptic complete integral of the first kind, $K'=K(k')$, and $k'$
is a complementary modulus, $k'^2=1-k^2$. The Hamiltonian $H$
obtained in this way then describes a pair of two
\emph{parity-even} associated Lam\'e systems, shifted one with
respect to the other for the half of the real period $2L$ of the
potential, that is equal to $2K$ for $C_m\neq C_l$, and $K$ for
$C_m=C_l$.  The cases with $C_mC_l= 0$ correspond to the Lam\'e
system \cite{Int}. By the Landen transformation \cite{Batem}, the
case $C_m=C_l$ is reduced to the case of the Lam\'e system with
same value of parameter $C_m$ but with $C_l=0$, and with the same
imaginary period but the real period divided in two. Taking  into
account that $C_{-m-1}=C_m$, without loss of generality one can
assume that $m>l\geq 0$. Isospectral subsystems $H^+_{m,l}$ and
$H^-_{m,l}$ belong to a class of finite-gap systems with even
potential and number of energy gaps in the spectrum equal to $m$
\cite{Int,Gang}.

\emph{Hidden bosonized supersymmetry -- } Consider an $n$-gap
periodic system with \emph{even} Hamiltonian
$H=-\frac{d^2}{dx^2}+V(x)$, $V(x+2L)=V(x)=V(-x)$. Its spectrum
$\sigma(H)$ is characterized by the band structure
$\sigma(H)=[E_{0},E_{1}]\cup\ldots\cup
[E_{2n-2},E_{2n-1}]\cup[E_{2n},\infty)$,  $E_0<E_1<\ldots
<E_{2n}$, which consists of $n$ valence bands and a conduction
band, separated by energy gaps corresponding to $n$ prohibited
bands. The $2n+1$ singlet band-edge states of definite parity and
energies $E_j$, $j=0,\ldots, 2n$, are given by periodic or
antiperiodic states of periods $2L$ or $4L$. The states in the
interior of permitted bands are described by Bloch-Floquet
quasi-periodic functions, and every internal energy level is
doubly degenerate. The energy doublets are distinguished by the
reflection (parity) operator $R$, $R\psi(x)=\psi(-x)$, that is a
nonlocal integral of motion. On the other hand, double
degeneration of energy levels is a characteristic feature of a
quantum mechanical $N=2$ supersymmetric system. The presence of
$2n+1$ singlets is an indication of the higher order ($\geq 3$)
nature of the corresponding hidden supersymmetry. Any finite-gap
system is characterized by  a nontrivial integral of motion
$A_{2n+1}$ that is a  self-conjugate differential operator of the
form
$A_{2n+1}=i\frac{d^{2n+1}}{dx^{2n+1}}+\alpha_{2n-1}(x)\frac{d^{2n-1}}{dx^{2n-1}}+\ldots
\alpha_0(x)$. The
 $(A_{2n+1},H)$ is the Lax pair of the $n$-th order Korteweg-de Vries (KdV) equation,
and the condition $[A_{2n+1},H]=0$ defines the stationary KdV
hierarchy. This pair of commuting operators satisfies identically
the relation \cite{Int}
\begin{equation}
    A_{2n+1}^2=P_{2n+1}(H),\quad P_{2n+1}(H)=\prod_{j=0}^{2n}(H-E_j),
    \label{spectral}
\end{equation}
and $2n+1$ singlet states $\Psi_j$ are the common eigenstates of
$H$ and $A_{2n+1}$ of the eigenvalues $E_j$ and $0$. The square of
the self-conjugate operator $A_{2n+1}$  is positive semi-definite,
and (\ref{spectral}) implies  the described band structure of the
spectrum. Integral  $A_{2n+1}$  is parity-odd, $\{R,A_{2n+1}\}=0$.
Taking into account that $[R,H]=0$ and $R^2=1$, we see that here
the reflection operator plays the same role as the operator
$\sigma_3$ for  $N=2$ superextended matrix system, and the
operator $Z=A_{2n+1}$ can be identified as the supercharge. Define
a nonlocal odd operator $\tilde{Z}=iRZ$. Odd supercharges $Q_1=Z$,
$Q_2=\tilde{Z}$ generate the $N=2$ nonlinear supersymmetry of
order $2n+1$:
\begin{equation}\label{susybos}
    \{Q_a,Q_b\}=2\delta_{ab}P_{2n+1}(H).
\end{equation}
Since this
structure appears in the one-dimensional system without matrix
(spin) degrees of freedom, the described nonlinear supersymmetry
of order $2n+1$ for any $n$-gap periodic system with parity-even
Hamiltonian is identified as a hidden \emph{bosonized }
supersymmetry \cite{SUSYbos,FM,hidLame}.

\emph{Supersymmetry and band structure -- }  Let us return to our
finite-gap self-isospectral system. A vector space spanned by
singlet band-edge states of the subsystem $H^+_{m,l}$ or
$H^-_{m,l}$ is divided into two vector subspaces formed by
$2L$-periodic and $2L$-antiperiodic (i.e., $4L$-periodic) states.
The singlet state $\Psi_0$ with the lowest energy $E_0$ is
$2L$-periodic, and the singlet state of the other edge of the
first valence band, $\Psi_1$, is antiperiodic. The two edge-states
$\Psi_{2j-1}$ and $\Psi_{2j}$, $j=1,\ldots,m$, separated by an
energy gap, have the same period \cite{Int,Gang}. Therefore, the
space of periodic singlet states has odd dimension, and the space
of antiperiodic states has nonzero even dimension. On these two
subspaces of singlet states, two irreducible non-unitary
representations of the $sl(2,\R)$ algebra are realized. Namely,
according to \cite{Gang,sl2}, the space of $2m+1$ singlet states
of the associated Lam\'e system with $m>l$ can be treated as a
direct sum of two $sl(2,\R)$-representations of dimensions $m-l$
(spin $j_1=\frac{1}{2}(m-l-1)$)  and $m+l+1$ (spin
$j_2=\frac{1}{2}(m+l)$). The period of the states of these
subspaces is dictated by the parity of $m-l$. When $m-l$ is odd,
spin-$j_1$ (spin-$j_2$) representation is realized on
$2L$-periodic ($4L$-periodic) states, for even $m-l$  the
periodicity of spin-$j_1$ and spin-$j_2$ subspaces interchanges.
Making use of the two corresponding algebraization schemes
\cite{sl2}, we find two commuting anti-diagonal self-conjugate
integrals of motion, $X$ and $Y$, where
\begin{equation}
    Y=i^{\epsilon_Y}\left(
    \begin{array}{cc}
    0 & Y_{m,l}^{-}(x) \\
    Y_{m,l}^{+}(x) & 0
    \end{array}%
    \right),
     \label{XY}
\end{equation}%
$
Y_{m,l}^{-}(x)=\frac{\mathrm{dn}^{m+1}\,x}{\mathrm{cn}^{m+l+2}\,x}\left(
\frac{\mathrm{cn}^{2}\,x}{\mathrm{dn}x}\frac{d}{dx}\right)
^{m+l+1}\frac{ \mathrm{dn}^{l}\,x}{\mathrm{cn}^{m+l}\,x},$
$Y_{m,l}^{+}(x)=Y_{m,l}^{-}(x+K), $ $\epsilon_Y=1$ $(0)$ for $m+l$
even (odd), and  $X$ has a form similar to (\ref{XY}) with
$X_{m,l}^{-}(x)=Y_{m,-l-1}^{-}(x)$,
$X_{m,l}^{+}(x)=X_{m,l}^{-}(x+K)$, $\epsilon_X=1-\epsilon_Y$. The
order of the differential operator $X$, $|X|=m-l$, is less than the
order of $Y$, $|Y|=|X|+2l+1=m+l+1$. $X$ and $Y$ have opposite
parities $(-1)^{m-l}$ ($X$), and $(-1)^{m-l+1}$ ($Y$). When $m-l$ is
odd, $2L$-periodic ($4L$-periodic) singlet states of each subsystem
are zero modes of the integral $X$ ($Y$). For even $m-l$ the role of
the operators $X$ and $Y$ as annihilators of periodic and
antiperiodic edge states interchanges.  The anticommutator
$\{X,Y\}=2Z$ produces a diagonal integral $Z=diag\,
(Z^+_{m.l},Z^-_{m,l})$, $|Z|=|X|+|Y|=2m+1$, with down,
\begin{equation}\label{Z}
    Z_{m,l}^{-}(x)=iY_{m,l}^{+}(x)X_{m,l}^{-}(x)=iX_{m,l}^{+}(x)Y_{m,l}^{-}(x),
\end{equation}
and up, $Z_{m,l}^{+}(x)=Z_{m,l}^{-}(x+K)$,  components, which are
the parity-odd integrals of motion annihilating all the singlet
states of the $m$-gap subsystems $H^-_{m,l}$ and $H^+_{m,l}$
described above. Hence, $Z^2=P_Z(H)$, where $P_Z(H)$ is a spectral
polynomial (\ref{spectral}) of order $2m+1$ with
$H=diag\,(H^+_{m,l},H^-_{m,l})$. {}From the explicit form of the
integrals $X$, $Y$ and $Z$ one finds that $[Z,X]$=$[Z,Y]=0$, and
$    X^2=P_X(H),$  $Y^2=P_Y(H)$.
The polynomials $P_X(H)$ and $P_Y(H)$ factorize the spectral
polynomial $P_Z(H)$, $P_Z(H)=P_X(H)P_Y(H)$, and include those
factors $(H-E_j)$ for which $E_j$'s are the eigenvalues of band-edge
states of corresponding periodicity \cite{ASfoot}.

Periodic (antiperiodic) states have an even (odd) number of nodes
in the period interval. The maximal number of nodes that can have
the band-edge states annihilated by $X^\pm_{m,l}$ and
$Y^\pm_{m,l}$ is not more than the order of these differential
operators. When $m-l$ is \emph{even}, $X$ annihilates $m-l$
anti-periodic band-edge states with $1,1,3,3,\ldots, m-l-1,m-l-1$
nodes. The $m+l+1$ periodic edge states  annihilated by $Y$ have
$0,2,2\ldots,(m+l),(m+l)$ nodes. We find that when $m-l$ is
\emph{odd}, $X$ annihilates $m-l$ periodic band-edge states with
$0,2,2,\ldots,m-l-1, m-l-1$ nodes if $m-l>1$, and one nodeless
state $\Psi^-_0$ if $m-l=1$. In the last case $X$ is the usual
first order supercharge \cite{const}. The operator $Y$ annihilates
$m+l+1$ anti-periodic states with $1,1,\ldots,m+l,m+l$ nodes.

\begin{figure}[tbp]
\includegraphics[width=0.95\linewidth]{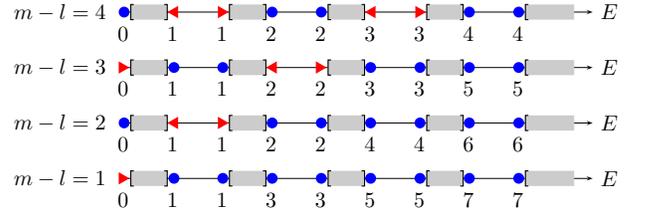}
\caption{(color online). Scheme of band structure for
self-isospectral systems with $m=4$. Triangles (dots) indicate
band-edge states annihilated by $X$ ($Y$), the digits below mean
their node numbers. The states with even (odd) number of nodes are
periodic (anti-periodic).} \label{fig:bands}
\end{figure}

The picture can be summarized as follows.  The band-edge state
$\Psi_0$ is a zero mode of the parity-odd supercharge, i.e. of $Y$
($X$) when $m-l$ is even (odd). The band-edge states of the same
prohibited band `attract' each other; they appear as zero modes of
the same supercharge. When the number of permitted bands $m+1$ is
fixed, and $m-l$ increases in steps of $2$, there appear two new
band-edge states annihilated by $X$, with increasing energies.
Every such pair is separated by a pair of zero modes of $Y$. The
highest $2(l+1)$ singlet states are zero modes of $Y$. These
properties are illustrated on Fig. \ref{fig:bands}.

Integral $Z$ reflects the degeneration of the states of each
subsystem, while  $X$ and $Y$ reveal the self-isospectrality of
the composed system. As a result, it is characterized by the
$4$-fold degeneration of quasi-periodic states and the double
degeneration of the band-edge states.

\emph{Nonlinear superalgebra -- } Besides nontrivial integrals
$X$, $Y$ and $Z$, our system is characterized also by mutually
commuting integrals $\Gamma_1=\sigma_3,$ $\Gamma_2=R$ and
$\Gamma_3=\sigma_3 R$. Any of them can be chosen as the operator
$\Gamma$ that classifies all the integrals into bosonic and
fermionic operators. Appropriate linear combinations of physical
states for which $\Gamma=+1$ and $-1$ are identified  as bosonic
and fermionic states. Let us choose $\Gamma=\sigma_3$. Eight
integrals $X$, $\Gamma_iX$, $Y$ and $\Gamma_iY$, $i=1,2,3$,
anticommuting with $\Gamma$ are identified as fermionic operators.
They anticommute between themselves for certain linear
combinations of the $8$ bosonic operators $Z$, $\Gamma_iZ$,
$\Gamma_i$ and $H$ with coefficients that are some polynomials in
$H$. Linear combinations of the bosonic operators ${\cal
J}_{1}^{(\pm) }=-\frac{i}{2}R\sigma _{3}Z\Pi _{\pm}$, ${\cal
J}_{2}^{(\pm)}=\frac{1}{2}\sigma _{3}Z\Pi _{\pm }$ and ${\cal
J}_{3}^{(\pm) }=- \frac{1}{2}R\Pi _{\pm },$ where $\Pi _{\pm
}=\frac{1}{2}\left( 1\pm \sigma_3 \right)$, have the only
nontrivial commutators
\begin{equation}
\label{su2}
    [{\cal J}^{(\pm)}_a,{\cal
    J}^{(\pm)}_b]=i\rho_c(H)
    \epsilon_{abc}{\cal J}^{(\pm)}_c.
\end{equation}
Here $a,b,c=1,2,3$,  $\rho_{1,2}=1$, $\rho_3=P_Z(H)$. This is a
nonlinear deformation of $su(2)\oplus su(2)\oplus u(1)\oplus
u(1)$, where the two last terms correspond to $\Gamma=\sigma_3$
and $H$. The nonlinear algebra (\ref{su2}) is reminiscent of
nonlinear symmetry algebra generated by the angular momentum and
Laplace-Runge-Lenz vector operators in the quantum Kepler problem
\cite{Pauli}. The complete superalgebra is identified as a
nonlinear deformation of the $su(2|2)$ superunitary symmetry, in
which $H$ plays a role of the multiplicative central charge
\cite{su22}.

\begin{figure}[tbp]
\includegraphics[width=1\linewidth]{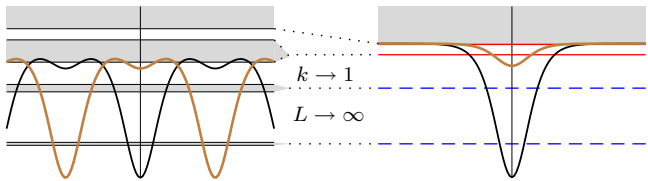}
\caption{(color online). Qualitative picture of supersymmetry
breaking in a self-isospectral system  with $m=3,$ $l=1$ in the
infinite-period limit. The form of potentials and bands shown on the
left corresponds to the modulus $k$ close to 1. Two lower horizontal
dashed lines on the right show energy levels of singlet bound
states, the upper separated horizontal continuous line corresponds
to a doublet of bound states, the line at the bottom of continuous
spectrum indicates a doublet of the lowest states of the scattering
sector.} \label{fig:limit}
\end{figure}

\emph{Infinite period limit -- } In the self-isospectral system
considered here, band-edge states form energy doublets,
quasi-periodic states are organized in quadruplets.
 In the infinite period limit, corresponding to $k\rightarrow
1$, $k'\rightarrow 0$, $K\rightarrow \infty$,
$\mathrm{dn}(x,k)\rightarrow \frac{1}{\mathrm{cosh}\, x}$, the
system transforms into a pair of reflectionless P\"oschl-Teller
systems given by potentials $V^\pm_{m,l}=-C^\pm \mathrm{cosh}^{-2}
x+c$, $C^+=C_l$, $C^-=C_m$. In this limit the spectral polynomial
degenerates \cite{FM}. The quasiperiodic states of the conduction
band reduce to the scattering states, supersymmetric doublet of its
band-edge reduces to a doublet of the lowest states of the
scattering sector. In a shifted subsystem $H^+_{m,l}$, $m-l$ lower
valence bands disappear, while the rest of them in both subsystems
shrink to the bound states. The resulting system is characterized by
$m-l$ singlet and $l$ doublet bound states, and by a doublet of the
lowest states of the scattering sector. The rest of the scattering
states is organized in energy quadruplets. This unusual picture of
supersymmetry breaking, illustrated on Fig. \ref{fig:limit}, is
related to the nonlinear nature of self-isospectral supersymmetry.

\emph{Conclusion -- } In the model investigated here the nonlinear
self-isospectral supersymmetry originates from: \emph{i}) the
separability of the singlet band-edge states of both subsystems
into two non-empty subspaces of periodic and anti-periodic states,
and \emph{ii}) the related factorization of the higher order Lax
pair operator of the associated stationary KdV hierarchy. The
mutually commuting integrals $X$ and $Y$ are the annihilators of
the band-edge states of definite periodicity. They factorize the
integral $Z$ that annihilates all the band-edge states. The
unusual nonlinear supersymmetry generated by these nontrivial
integrals together with integral $\sigma_3$ and parity operator
$R$, reveals the band structure of the system and all its
peculiarities in the same way as a the nonlinear symmetry
associated with the Laplace-Runge-Lenz vector reflects specific
properties of the hydrogen atom spectrum \cite{Pauli}.

We thank B. Dubrovin, V. Enolskii, I. Krichever, A. Treibich, R.
Weikard and A. Zabrodin for valuable communications. The work was
supported by FONDECYT (Grants 1050001, 7070024 and 3085013), CONICYT
and DICYT (USACH), by M\v SMT  \v CR project Nr. LC06002, and by
projects MTM2005-09183 and VA013C05. LMN thanks Universidad de
Santiago de Chile for hospitality.

\end{document}